\documentclass[mypaper,8pt,twoside]{CoAst}
\usepackage{epsf,graphicx,fancyhdr}
\renewcommand{\chaptermark}[1]{\markboth{#1}{}}

\newcommand{\kopf}{\small\itshape Comm. in Asteroseismology \\ Contribution to the Proceedings of the Wroclaw HELAS Workshop, 2008}

\newcommand{\Authors}[1]{\begin{center}\normalsize\bf\sf #1 \end{center}}

\renewcommand{\author}[1]{\begin{center}\normalsize\bf\sf #1 \end{center}}
\newcommand{\Address}[1]{\begin{center}\small\sf #1 \end{center}}

\newcommand{\References}[1]{\begin{flushleft}{\large References\\}\vspace*{2mm}\small #1 \end{flushleft}}

\newcommand{\chapterCoAst}[2]{\chapter[\sf\normalsize #1\\ \footnotesize \hspace*{5mm}by #2 \sf\normalsize][]{#1\\}\rhead[\fancyplain{}{\sf\footnotesize \center{#1}}]{\fancyplain{}{\sffamily\thepage}}\lhead[\fancyplain{\kopf}{\sffamily\thepage}]{\fancyplain{\kopf}{\sf\footnotesize \center{#2}}}}

\newcommand{\figureDSSN}[5]{\begin{figure}[#4]
\centering
\includegraphics*[#5]{#1}
\caption{#2}
\label{#3}
\end{figure}}

\renewcommand{\chaptername}{}

\newcommand{\acknowledgments}[1]{\vspace*{5mm}\noindent  \textbf{Acknowledgments.} #1}

\def\rfr{\smallskip\par\noindent
        \hangindent=7truemm
        \hangafter=1}

\begin{document}
\sf

\chapterCoAst{Can opacity changes help to reproduce the hybrid star pulsations?}
{T. Zdravkov and A. A. Pamyatnykh} 
\Authors{T. Zdravkov$^{1}$ \& A. A. Pamyatnykh$^{1,2}$} \Address{
$^1$ N. Copernicus Astronomical Center, ul. Bartycka 18, 00-716 Warsaw, Poland\\
$^2$ Institute of Astronomy, Russian Academy of Science, Pyatniskaya
Str. 48, 109017 Moscow, Russia\\
}

\noindent Hybrid stars like $\nu\,$Eri and $12\,$Lac show two different types of pulsations: (i) low-order acoustic and gravity modes of $\beta\,$Cephei type with periods of about $3-6$ hours, and (ii) high-order gravity modes of the SPB type with periods of about
$1.5-3$ days. Seismic models of $\nu\,$Eri (Pamyatnykh et al. 2004, Dziembowski \& Pamyatnykh 2008) using both OPAL (Iglesias \& Rogers 1996) and OP (Seaton 2005) opacity data well reproduce the observed range of short period low-order pulsations of the $\beta\,$Cep type and also show a tendency to instability of long period high-order gravity modes. With the OP data, the instability of the quadrupole
($\ell=2$) high-order gravity modes was found, but at slightly shorter periods than those observed. Trying to reproduce both short
and long period ranges of the $\nu\,$Eri pulsations, we tested the effects of artificial 50 \% opacity enhancement around the metal
opacity bump at $\log T\approx 5.3-5.5$, as it is shown on Fig$.\,1$.
\figureDSSN{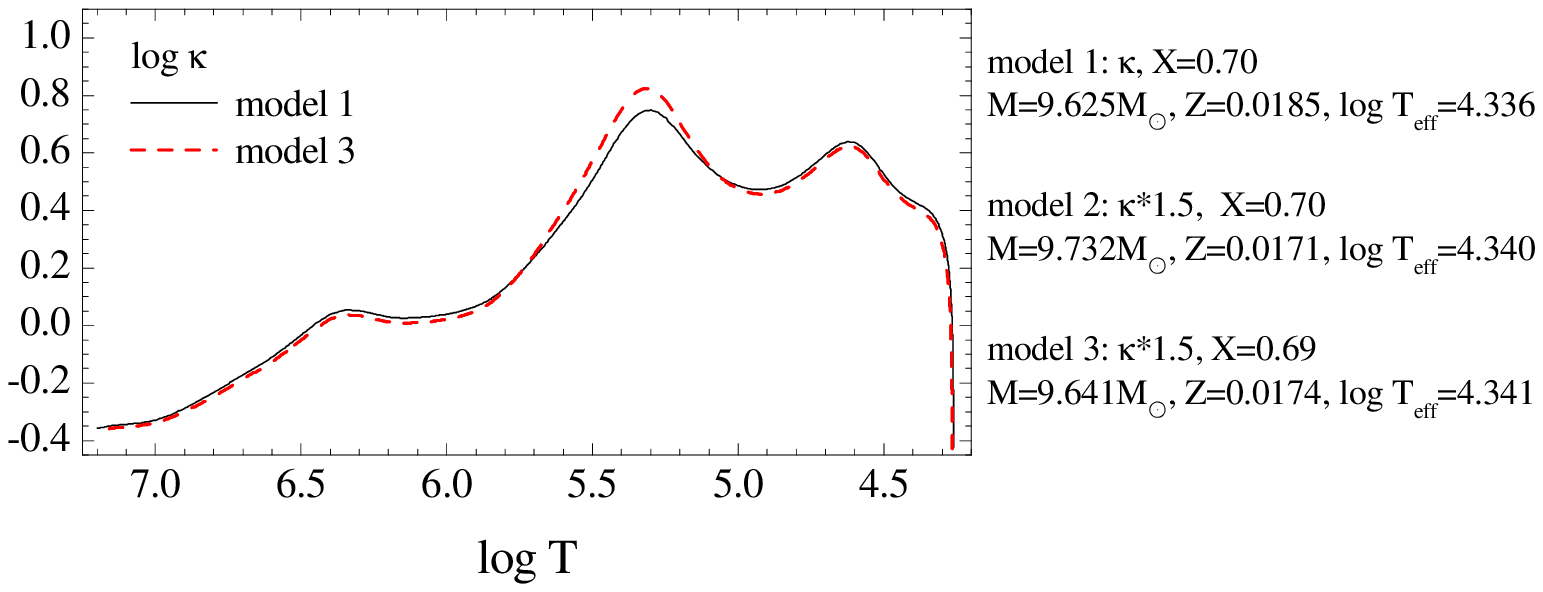}{Opacity behavior inside non-modified ($1$) and modified models ($2$, $3$). Model$\,2$ follows the line of Model$\,3$, for clearance it was not depicted.}{kappa run}{!h}{clip,angle=0,width=110mm}
Models were computed using the OP opacities and new solar proportions in the heavy element abundances (A04, see Asplund et al. 2005). In all models, the frequencies of radial fundamental and two dipole modes ($g_{1}$ and $p_{1}$) were fitted to the observed values with accuracy better than $0.0005$ c/d. As we may see, in Fig.\,1, the opacity enhancement in the modified models is less than 50 \%, because the fitted models differ in metal abundances and other parameters.

As a result of the opacity enhancements, Model$\,3$ is very close to have unstable dipole high order gravity modes at the observed frequency range (Fig.\,2).
\figureDSSN{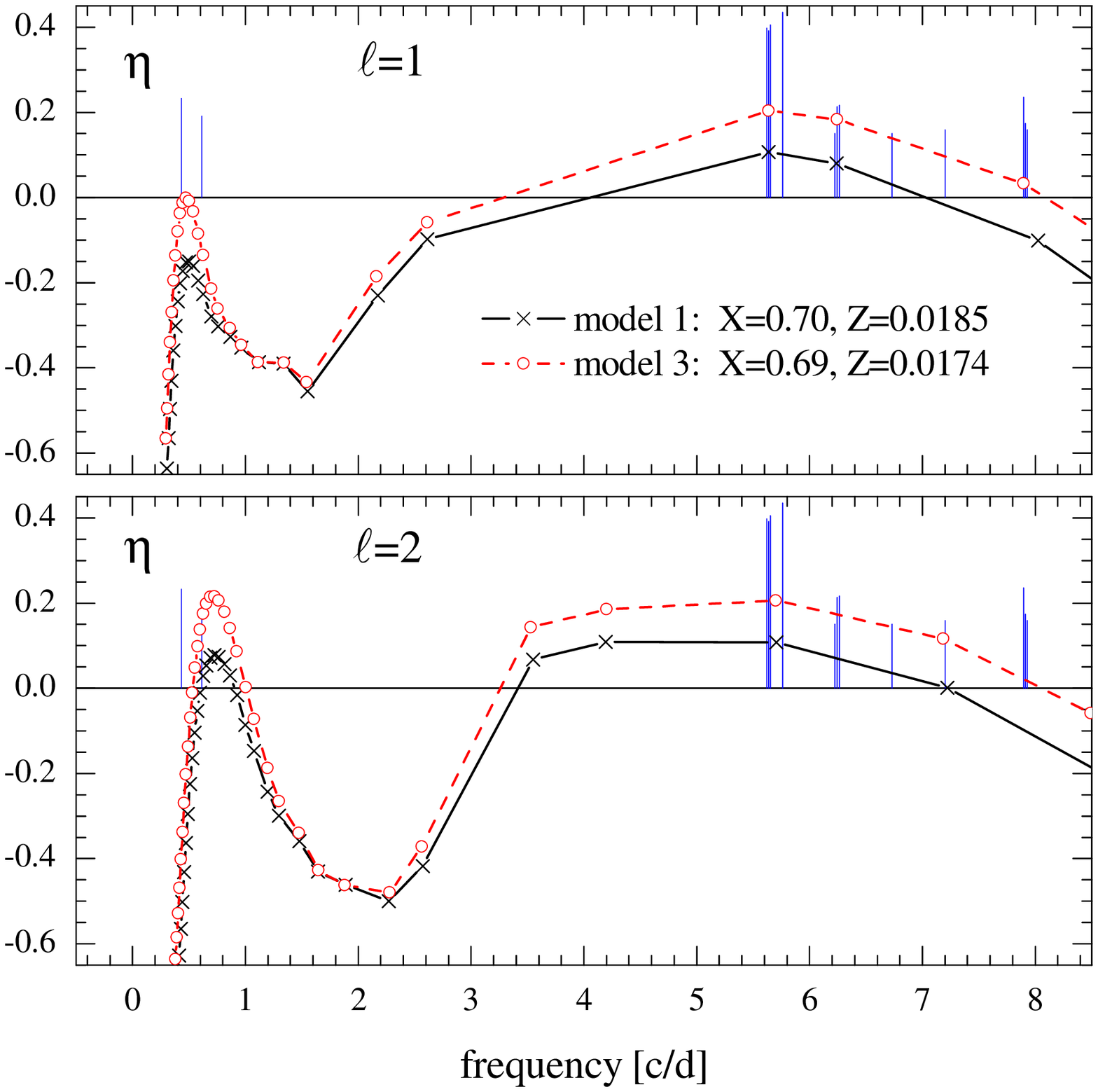}{The normalized growth rates, $\eta$, of $\ell=1$ and $\ell=2$ modes as a function
of mode frequency in seismic models of $\nu\,$Eri ($\eta>0$ for unstable modes). Vertical lines mark the observed frequencies
(Jerzykiewicz et al. 2005), with amplitudes given in a logarithmic scale. In all models, the frequencies of radial fundamental and two
dipole modes ($g_{1}$ and $p_{1}$) fit the observed values at $5.763$, $5.637$ and $6.244$ c/d, respectively. In Model 3, one more
dipole mode ($p_{2}$) fits the observed value at $7.898$ c/d (with the accuracy $0.001$ c/d). Model$\,2$  nearly follows the line of the Model$\,3$, for clearance it was not depicted.} {growth rate}{!h}{clip,angle=0,width=92mm}
Also, the range of the unstable short period modes is in a better agreement with the observations. In addition, a small decrease of the hydrogen abundance ($X=0.69$ instead of $X=0.70$) allows to achieve also a good frequency fit of the $\ell=1$, $p_{2}$ mode to the observed value of $7.898$ c/d.

We note however that the required opacity increase, by $50\,\%$ in the Z bump, appears larger than allowed by uncertainties in current opacity calculations. A similar improvement in the fit may be achieved with a more modest (few percent) modification of the opacity bump at log$\,T=6.3$. We plan to examine this option in future.

\acknowledgments{We acknowledge partial financial support from the Polish MNSiW grant No. 1 P03D 021 28 and from the HELAS project.}

\References{ \rfr Asplund N., Grevesse N., Sauval A. J., 2005, in Barnes III T. G., Bash F. N., eds, ASP Conf. Ser. vol. 336, The Solar Chemical Composition, p.25 \rfr Dziembowski W.A. \& Pamyatnykh A. A., 2008, MNRAS, 385, 2061 \rfr Iglesias C. A. \& Rogers F. J., 1996, ApJ, 464, 943 \rfr Jerzykiewicz M., Handler G., Shobbrook R. R., et al., 2005, MNRAS, 360, 619 \rfr Pamyatnykh A. A., Handler G. \& Dziembowski W. A., 2004, MNRAS, 350, 1022 \rfr Seaton M. J., 2005, MNRAS, 362, L1 }

\end{document}